\documentclass[lettersize,journal]{IEEEtran}

\usepackage{todonotes}
\usepackage{color,soul}
\usepackage{threeparttable}
\usepackage{multirow}
\usepackage{graphicx}
\usepackage{float}
\usepackage{todonotes}
\usepackage{array}
\usepackage{subcaption}
\usepackage{booktabs}
\usepackage{amsmath}
\usepackage{algorithm}
\usepackage[noend]{algorithmic}
\usepackage{tabularx}
\usepackage{xfrac}
\usepackage{textcomp}
\usepackage{mathtools}
\usepackage{xcolor}
\usepackage{enumitem}

\usepackage{xr}

\newcolumntype{Y}{>{\centering\arraybackslash}X}

\usepackage{hyperref}

\begin{document}

\title{CIMPool: Scalable Neural Network Acceleration for Compute-In-Memory using Weight Pools}

\author{Shurui Li, and Puneet Gupta,~\IEEEmembership{Fellow,~IEEE}}



\maketitle

\begin{abstract}
Compute-in-memory (CIM) based neural network accelerators offer a promising solution to the Von Neumann bottleneck by computing directly within memory arrays. However, SRAM CIM faces limitations in executing larger models due to its cell size and on-chip memory constraints. This work proposes CIMPool, a CIM-aware compression and acceleration framework that counters this limitation through a weight sharing-based compression technique, aptly named `Weight Pool,' enabling significantly larger neural networks to be accommodated within on-chip memory constraints.
This method minimizes the accuracy trade-off typically associated with parameter compression, allowing CIMPool to achieve a significantly larger compression ratio compared to the traditional quantization method with iso-accuracy.

Furthermore, CIMPool co-optimizes the compression algorithm, hardware, and dataflow to efficiently implement the hardware permutation required by weight pool compression, with negligible area and throughput overhead. Empirical results demonstrate that CIMPool can achieve 8-bit level accuracy with an effective 0.5-bit precision, reduce chip area by 62.3\% for ResNet-18, and enable the execution of an order of magnitude larger models for a given area budget in SRAM CIMs. When DRAM is used to store weights, CIMPool can reduce the total energy by $3.24 \times$ compared to iso-accuracy traditional CIMs.
\end{abstract}




\section{Introduction}
Compute-in-memory (CIM) based neural network accelerators have proliferated in recent years because of their unique advantage in alleviating the Von Neumann bottleneck. \cite{jhang2021cimtrends, sehgal2021cimtrends}. This is achieved by computing directly within memory arrays, significantly reducing data movement overhead. \cite{wan2022compute, ali2022compute, yu2020compute, jain2017computing}
Among various CIM accelerators, SRAM CIM is a promising candidate because of its technological maturity, high reliability, and low latency. Additionally, SRAM's low write latency and high endurance enable the weights stored in the CIM to be rewritable at runtime, enhancing flexibility.
However, SRAM CIM faces its own set of challenges. Its relatively large cell size makes area a limiting factor for most SRAM CIM accelerators, often hindering the execution of larger models. With the continual increase in the size of modern neural networks, improving the area efficiency of SRAM CIM is crucial to facilitate the execution of these larger networks.

Due to the area limitations of SRAM CIM, a model that fits entirely on the CIM is typically too small to be useful without any additional compression techniques. Even with aggressive scaling, a large SRAM capacity is 16MB, which can at most contain 16M weights quantized to 8-bit. For example, this may hold an 8-bit ResNet18 but is too small to contain ResNet34 or larger. \cite{yu2021trends} On the other hand, if the network is too large to be contained solely on the CIM, additional performance costs are incurred to move the weights from the DRAM. Thus, if the neural network weights are not made to fit into a small area, the inference is slower and more energy-hungry due to the additional latency (and energy) costs of loading the weights from DRAM. Especially for low-batch size inference regimes, this wipes out the performance benefits created by CIM in the first place, namely the reduced data movement.

While technology scaling and circuit-level optimizations are conventional solutions to this problem, our work adopts a novel approach, employing software and algorithm-level optimizations with co-optimization of CIM hardware. Traditional methods like quantization and compression effectively reduce the storage demands of neural networks but often overlook the underlying hardware. This oversight can lead to various overheads when implemented on CIMs, and some even make the implementation infeasible. To address these issues, we introduce CIMPool, a hardware-aware compression and acceleration framework tailored for scalable and efficient neural network acceleration in CIMs.

The proposed framework is based on the weight pool concept \cite{li2022weightpoolmlsys}, a compression method through weight sharing, which compresses neural networks by converting their weights into a small pool (codebook) of weight vectors and the indices to the weight pool. This compression method fits CIM well because the codebook is fixed and is typically small enough to be stored within a single CIM array, which means the CIM array content can be fixed for the entire execution of the neural network.

While the combination of weight pool and CIM seems promising, there are some challenges when applying weight pool to CIMs. One challenge is that the vector size of the original weight pool is just 8 for accuracy and implementation considerations, which is too small for CIM arrays that typically have a height a few times larger than 8, making the execution inefficient. However, naively increasing the vector size leads to a dramatic reduction in accuracy. To address this, CIMPool introduces a 1-bit error to mitigate the accuracy drop associated with the large vector size and further prunes the error to minimize the hardware overhead.

Another main impediment to directly applying the weight pool compression method to CIM is the requirement of hardware permutation. In weight pool compression, adjacent weight vectors are assigned to different weight pool vectors (CIM columns). However, since the weight pool of the CIM is fixed, the CIM outputs need to be permuted back to their correct order, which can lead to significant hardware and latency overhead. CIMPool addresses this issue by leveraging SRAM CIM's bit-serial processing property and co-optimizing the weight pool and hardware design.

The main contributions of this work can be summarized as follows:

\begin{itemize}[nosep]
    \item CIMPool compresses neural networks with underlying CIM in mind. Only two CIM arrays are required for processing an entire neural network.
    \item By leveraging bit-serial computation and co-optimization of dataflow and weight pool, CIMPool can permute the CIM outputs in hardware with negligible overhead. 
    \item For the Food-101 dataset, CIMPool can achieve 8-bit level accuracy with 27.7 $\times$ compression.
    \item CIMPool can reduce chip area by 62.3\% compared to iso-accuracy 4-bit CIMs. For a given area budget such as $100 mm^2$, CIMPool makes the execution of an order of magnitude larger models possible for SRAM CIMs.
    \item When DRAM is used to store the weights, CIMPool can reduce the total energy by $3.35 \times$ compared to iso-accuracy 4-bit CIMs. 
\end{itemize}


\section{Background and Motivation}
\subsection{SRAM CIM}
SRAM CIM emerges as a promising solution for energy-efficient neural network acceleration, merging computation and memory seamlessly to minimize both computation and data movement costs. An SRAM CIM array can perform a large vector-matrix multiplication in a single cycle, which is essential in neural network computations. In this setup, the neural network weights are mapped onto SRAM cells, while inputs are applied in parallel across the word lines. As inputs traverse through these word lines, each SRAM cell modulates the corresponding bit line's current, effectively executing the multiplication. ADCs (Analog-to-digital converters) at the end of bit lines collect these currents, representing dot product results, and convert them into digital signals. Due to the single-bit nature of SRAM cells, multi-bit weights are stored across multiple bitlines. Multi-bit inputs are processed bit-serially, with partial sums accumulated through analog or digital shift-and-add units. In mapping a neural network layer to a CIM array, output channels (or filters in CNNs) are typically mapped to CIM columns, while input channels are mapped to rows.

In CIMs, energy and area are usually dominated by the ADCs. To amortize the ADC overhead, therefore, most CIMs tend to use long bitlines (64-1024 sized). To maximize input reuse in dense CIM arrays, the wordlines also tend to be long, delivering a high degree of input reuse and parallelism. Though peak efficiency and throughput can be very high for such CIMs, this architecture also limits viable model compression techniques. For example, network pruning, if used, needs to be very structured due to the long bitlines and wordlines. Weight quantization (by the provisioned bit-parallel weight storage and execution in SRAM CIMs) and output quantization range (determined by the ADC) are essentially baked into the hardware irrespective of the neural network or the layers contained in them.

\subsection{Weight Pool Compression} \label{sec:weightpoolbackground}
Weight pool, a method for neural network compression through weight sharing, involves a small pool (or codebook) of weight vectors shared across the entire model. This allows weights to be replaced with indices to the weight pool, facilitating compression. Typically, the pool size (the number of weight vectors in the pool) is 256 or fewer, ensuring indices require only 8 bits or less. For instance, if the weight pool has a vector size of 8, each group of 8 weight values in the original model can be replaced by a single 8-bit index, resulting in a 32$\times$ compression compared to the floating-point baseline. A common approach to generating the weight pool is by performing k-means clustering on the pre-trained weights, using cluster centers as the pool content. 

The weight pool method holds an intriguing synergy with CIM - the fixed weight vectors in the pool could be housed in a single CIM array, shared across the entire neural network. This concept could potentially enable the execution of a complete model using just one CIM array. However, there are three primary challenges. First, these compression methods allow weight pool vectors to be reused by different filters without constraints, leading to significant underutilization and scheduling challenges in CIM deployments. Secondly, existing weight pool compression methods often use small vector sizes, like 8 \cite{li2022weightpoolmlsys} or 9 \cite{son2018weightpooleccv}, for accuracy reasons. These sizes are considerably smaller than a typical neural network layer's full dot product size. Consequently, the CIM arrays would have poor utilization. Moreover, a single dot product must be divided into multiple partial sums and accumulated over several cycles, which reduces throughput and could suffer more from ADC quantization. Thirdly, the CIM implementation of weight pool compression requires permuting the CIM outputs back to their original order in hardware, which will be discussed in Section \ref{sec:hwpermutation}.

\subsection{Related Work}
Prior research on CIM-based neural network accelerators has primarily focused on addressing two critical challenges: memory bottleneck and energy efficiency. The memory bottleneck issue arises mainly due to limited on-chip memory capacities, restricting the scale and complexity of neural networks deployable on CIM architectures. Concurrently, energy efficiency is pursued rigorously to optimize the trade-off between computational accuracy and power consumption in neural networks.

Existing approaches to alleviate memory constraints and enhance energy efficiency generally revolve around model compression techniques. Among these, structured pruning and tensor-train decomposition have become especially popular methodologies. Typically, these approaches aim to reduce model sizes significantly while maintaining acceptable performance levels.

Sie et al. \cite{sie2022mars} introduced a CIM-aware sparse neural network compression algorithm, achieving substantial memory savings. However, when compared to our work, CIMPool, their compression ratio is considerably lower. For example, under similar accuracy conditions, CIMPool achieves a 14.8× higher compression ratio than their approach (with only a 1

Yue et al. have contributed significantly to the CIM domain with various architectures. One of their designs utilizes a specialized CIM CNN processor optimized for sparse computation \cite{yue2022binarycim}. Despite demonstrating notable advancements, their processor relies on complex hardware implementations and only supports bitwidth configurations in powers of two, limiting flexibility compared to the CIMPool framework proposed in this paper.

Another relevant work by Chang et al. \cite{chang2022structured} employed structured pruning and mixed bit-width compression in CIM accelerators, efficiently addressing the energy-memory trade-off. However, their methodology requires extra controller logic and hardware overhead for maintaining bit-width tables, adding complexity and cost implications.

Yue et al. proposed both a CIM CNN processor leveraging bit-slice architectures \cite{yue2022cimcnn} and a binary-weight CIM processor \cite{yue2022binarycim}, optimizing energy consumption through selective ADC powering and specialized hardware. Although effective in their respective contexts, such solutions introduce additional complexity and specialized hardware constraints not present in CIMPool.

In contrast, the CIMPool method introduced in this work addresses these limitations by adopting a novel weight-pool compression technique, combined with algorithm and hardware co-design for optimal efficiency. CIMPool achieves significantly higher compression ratios while maintaining competitive accuracy levels and avoiding the need for highly specialized hardware support. Its efficiency and simplicity provide distinct advantages in practical implementations of CIM accelerators, positioning it favorably within existing literature on model compression for CIM architectures.

This paper builds upon and advances previous efforts, presenting CIMPool as a promising and broadly applicable solution to the area and energy challenges prevalent in CIM-based neural network accelerators.

\section{CIMPool algorithm} \label{sec: method}

Inspired by the properties of weight pool, we propose CIMPool to realize efficient deployment of weight pool on SRAM CIM. CIMPool is designed from the hardware perspective, which ensures it will not cause underutilization of CIM arrays or require significant scheduling overhead. However, as briefly discussed in the introduction, many modifications and optimizations are necessary for the efficient implementation of weight pool on SRAM CIMs.
In this section, we first introduce the high-level idea of CIMPool and then discuss the modifications and optimizations over conventional weight pool compression, followed by the overall flow of the CIMPool framework.

The overall idea of CIMPool is to hardcode the weight pool into one CIM array and share the single array with the entire neural network during execution, such that the computing part of the chip is simply a single SRAM CIM array, and the rest of the chip area can be allocated to weight and activation memory. Therefore, we set the weight pool's vector size and pool size to $128 \times 128$, respectively, a common SRAM array size, so that the weight pool can be stored in a conventional SRAM array without underutilization. The vector size and pool size can be configured for other array sizes as well. For the rest of the paper, the $128 \times 128$ CIM array is assumed for consistency. We also want to note that CIMPool, in theory, is also applicable to other array sizes without significant changes in the algorithm.

\begin{figure}[htp]
\centering
\includegraphics[width=0.9\columnwidth]{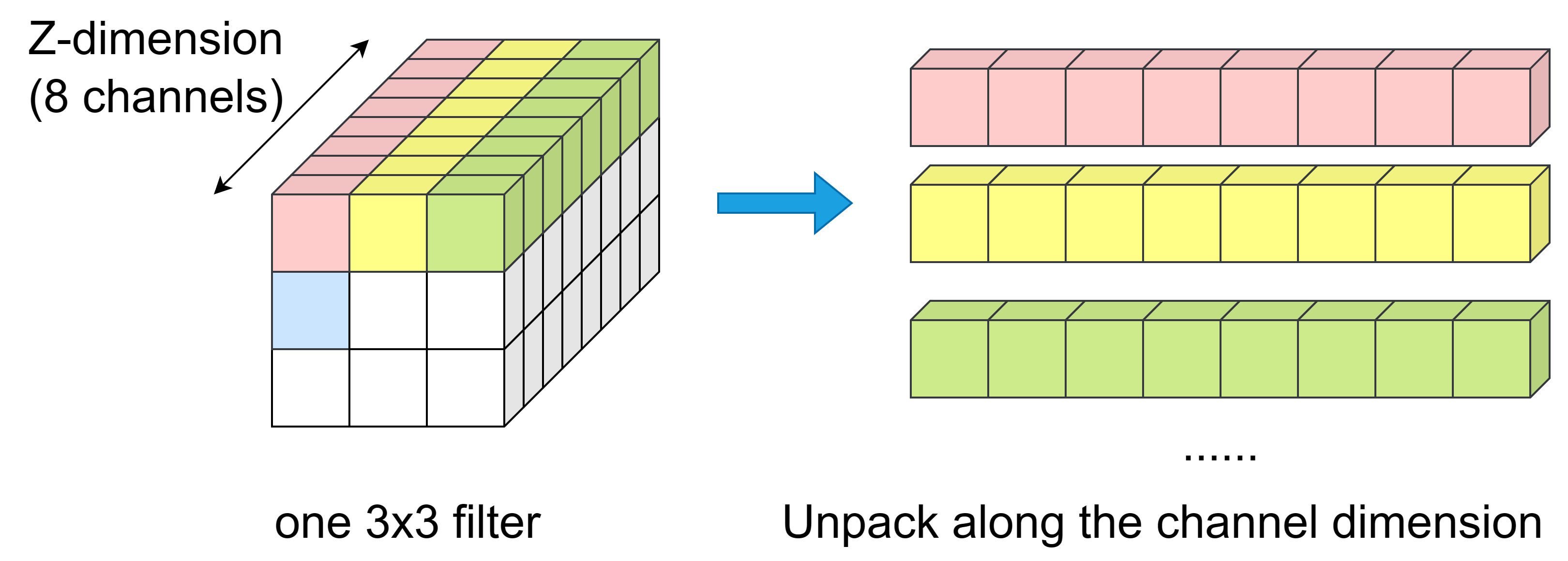}
\caption{Visualization of z-dimension weight packing.}
\label{fig:zdimgrouping}
\end{figure}

When assigning the weights to the vectors in the weight pool, CIMPool groups weights into vectors of 128 in the Z-dimension (channel dimension), as shown in Figure \ref{fig:zdimgrouping}, for better flexibility. Z-dimension weight grouping works for convolution layers with different filter sizes and is also applicable to dense layers.
By setting the vector size to 128, the compression ratio improves by $16\times$ compared to the original weight pool with a vector size of 8 - at the cost of a significant accuracy drop. To compensate for the accuracy drop while making the CIM hardware as efficient as possible, CIMPool introduces a series of modifications and optimizations, including the error term, error term quantization and pruning, error scaling factor, and weight pool quantization, which will be discussed in detail in this section.

\begin{figure*}[ht]
\centering
\includegraphics[width=0.9\linewidth]{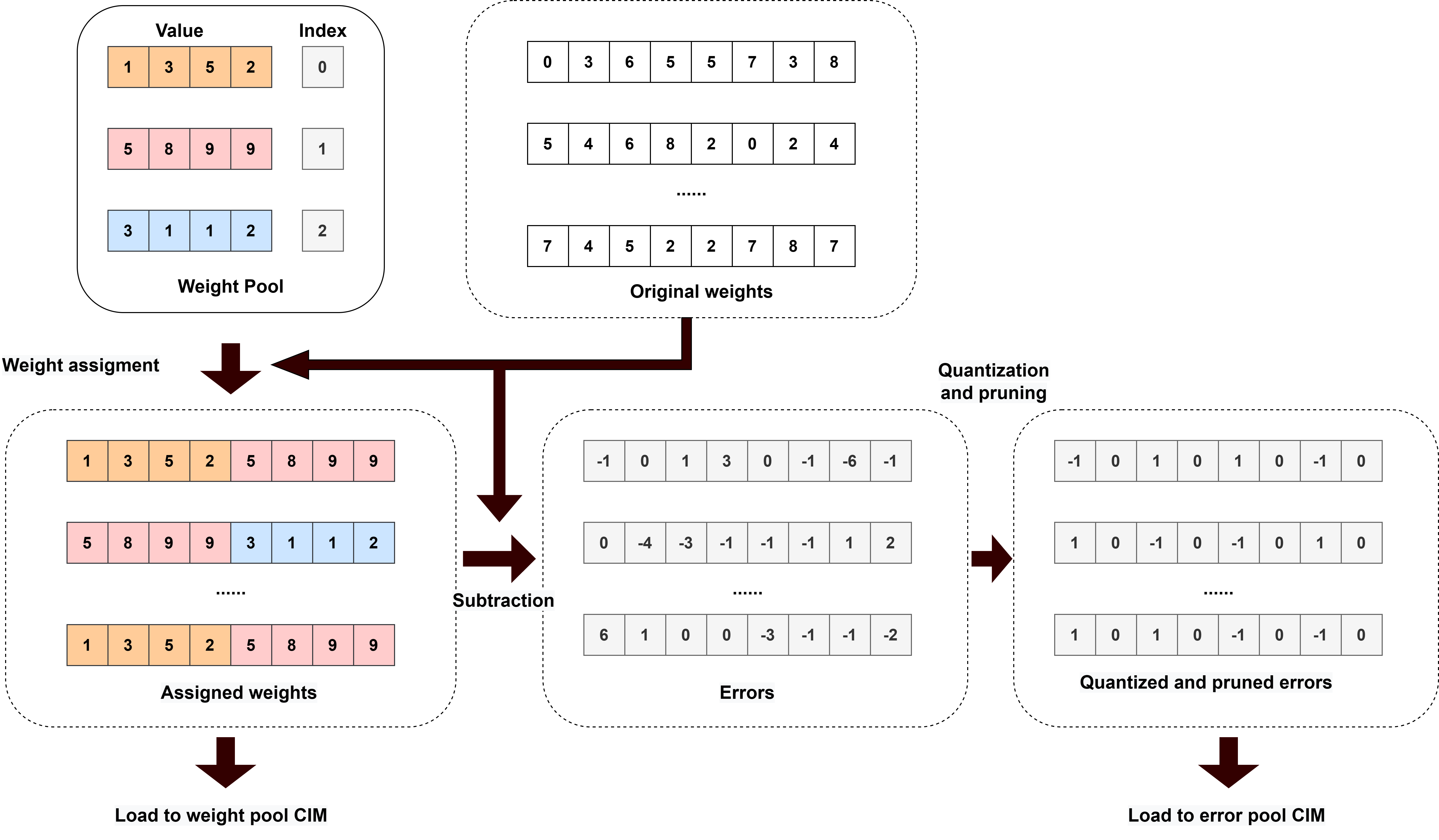}
\caption{Visualization of weight pool compression.}
\label{fig:weightpooloverview}
\end{figure*}

Figure \ref{fig:weightpooloverview} illustrates the compression process of CIMPool with a simplified weight pool that contains 3 vectors with size $1\times 4$. The original weights are first split into multiple weight vectors with shape $1\times 4$ and then their values are replaced with the closest weight vector in the weight pool. Therefore, indices of the weight pool are used to represent the weights instead of values to achieve compression. To compensate for the accuracy drop, the errors of the weight pool compression are computed and then further binarized and pruned to minimize the overhead. Details about the error term will be discussed later in this section.

\subsection{Non-repeating weight pool}

In the original weight pool compression, each vector from the original weights can freely choose the closest (most similar according to the given metric) vector from the weight pool, which means adjacent weight vectors from the original weights can select the same weight vector from the weight pool. While this works if the weight pool is used as a pure compression algorithm, it can cause significant conflicts and scheduling overhead when applied to CIM. For example, assuming a worst-case scenario where all 128 weight vectors that need to be scheduled on the CIM array are assigned to the same weight vector, conflicts occur, and only one of them can be scheduled to the CIM, causing significant underutilization. This could lead to $128\times$ worse throughput. While this is an extreme case, since the weight pool size is the same as the number of columns in the CIM array, such conflicts are common, and CIM array utilization can never reach 100\%. To prevent such underutilization, CIMPool constrains the weight vector assignment process so that no conflicts will occur.

To achieve this, the original weights are grouped into groups of $N$ vectors, where $N$ is the number of CIM columns, and all vectors in each group will be scheduled to the CIM at the same time. During the weight assignment process, a greedy algorithm is implemented such that each weight vector is assigned to a unique vector in the weight pool. With this constraint, CIMPool achieves 100\% array utilization.

\subsection{Error term}

\begin{figure}[htp]
\centering
\includegraphics[width=\columnwidth]{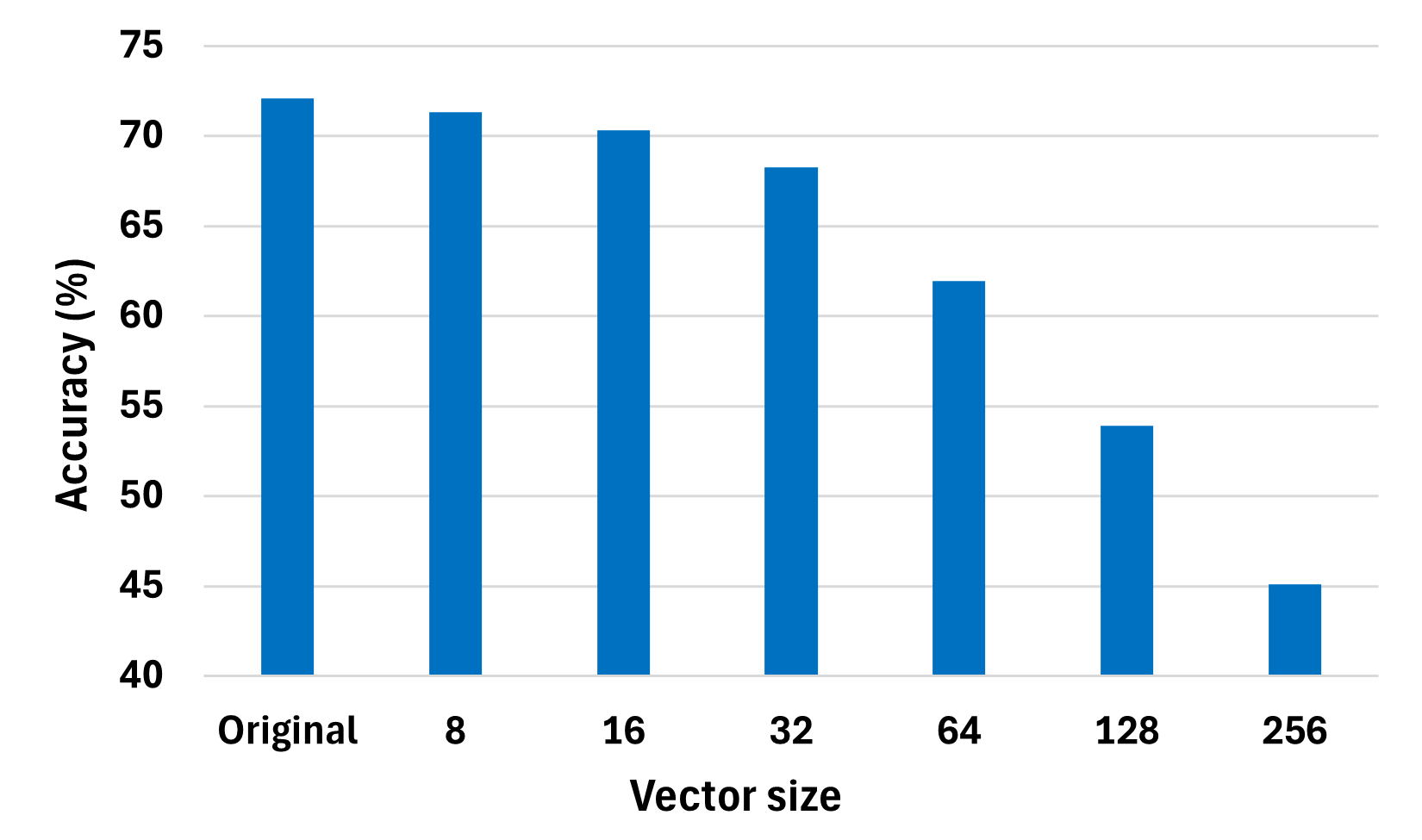}
\caption{Accuracy vs. vector size of weight pool compression, for ResNet-18 on CIFAR-100.}
\label{fig:accvsvecsize}
\end{figure}

Although setting the vector pool dimension to $(128,128)$ is a good fit for SRAM CIM, an issue with setting the vector size to a large number like 128 is that the accuracy will severely degrade, as shown in Figure \ref{fig:accvsvecsize}. This trend makes sense because the compression ratio goes up as the vector size increases, and there are significantly fewer unique permutations for the entire weight of a given layer. For example, if the vector size is 1, then each weight can choose from 128 values, which is similar to conventional 7-bit quantization, with good accuracy but almost no compression. Having a vector size of 128 means every 128 weights can only select from 128 unique vectors, which has a compression ratio comparable to 1-bit quantization and leads to a considerable accuracy drop. Figure \ref{fig:accvsvecsize} shows the accuracy versus vector size results with ResNet-18 on CIFAR-100 \cite{cifar}. From the plot, it's clear that accuracy degrades as vector size increases, and the accuracy drop is more than 15\% for vector sizes larger than 128.

To compensate for the accuracy degradation caused by the large vector size, we introduce the error term ($E$), which is a key concept in CIMPool. The error term $E$ is defined as:
\begin{equation}
    E = W_{ori} - W_{wp}
\end{equation}
, where $W_{ori}$ stands for the original weights and $W_{wp}$ stands for the weights after weight pool compression ($W_{wp}$ only consists of vectors in the weight pool). If the error term has high precision, the original weight can be fully reconstructed by adding the error term to $W_{wp}$. However, since in the baseline setup, every weight has a unique error term, the error term will result in significant storage and area overhead if a high bitwidth is used. Thus, in CIMPool, the is heavily quantized to reduce the overhead. In this case, the reconstructed weight $W_{rc}$ is defined as:
\begin{equation}
    W_{rc} = W_{wp} + E_{q}
\end{equation}
, where $E_{q}$ is the quantized error term.
During CIMPool execution, $W_{wp}$ is mapped to the weight pool array and $E_q$ is mapped to the error array. The outputs are computed as:
\begin{equation}
    O = I \times W_{wp} + I \times E_q
\end{equation}
, where $O$ stands for outputs and $I$ stands for inputs. This equation is equivalent to $I \times W_{rc}$

\paragraph{Error term quantization}

In practice, we find that with some tuning, even using a 1-bit error term is sufficient for most cases, achieving close to floating point accuracy, including some challenging datasets. Adjustment of the relative value of error is necessary to maintain accuracy, since directly setting errors to $+/- 1$ could make the quantized error significantly larger than the actual error and lead to poor accuracy. To address this, we profiled the network to obtain the mean absolute error for each layer after weight-pool compression. Then, we multiplied the quantized 1-bit error with the profiled mean absolute value. This tuning enables 1-bit error quantization to achieve accuracy on par with original accuracy.
Thus, CIMPool sets the bitwidth of error terms to 1 bit to minimize the area and storage requirements.

\subsection{Weight pool content and quantization}

With the newly introduced 1-bit error term, the requirement for the weight pool content and precision is greatly relaxed. With some tuning and optimization, the precision of the weight pool can be reduced to just 1 bit, effectively binarizing the weight pool. We empirically find that a randomly generated binarized weight pool can achieve the same level of accuracy compared to an 8-bit weight pool generated using K-Means clustering of the pretrained weights when paired with a 1-bit error term. Therefore, in CIMPool, the weight pool content is randomly generated binarized values (1 and -1). Similar to the binary error term, the binary weight pool is scaled with the mean absolute value of the original weights.

We want to emphasize that this observation does not mean that the weight pool is superfluous. In CIMPool, the weight pool plays a pivotal role as the error term is not a direct trainable parameter but is determined based on the weight pool content. The effectiveness of the weight pool is demonstrated by the significantly higher accuracy when compared to binary weight networks (which have the same compression ratio as default CIMPool), which is presented in Section \ref{sec:eval}, as well as the accuracy difference of different weight pool sizes that are discussed in Section \ref{sec:hwpermutation}.

\subsection{Error pruning}

\begin{figure*}[htp]
\centering
\includegraphics[width=0.8\linewidth]{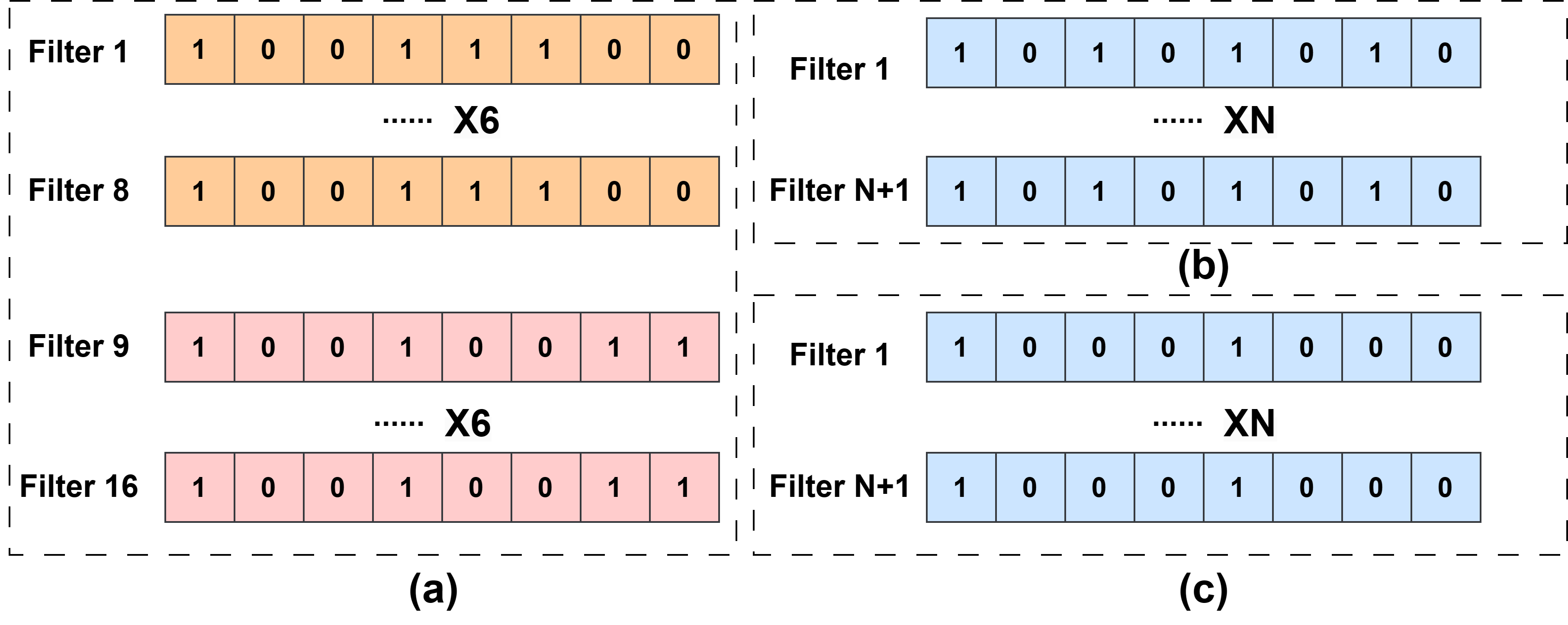}
\caption{(a) Zero masks of conventional semi-structured pruning, zero positions within the mask are not structured, but the mask is shared with a group of vectors to reduce storage overhead. (b), (c): Zero masks of our structured pruning approach with 50\% and 75\% sparsity.}
\label{fig:errorpruning}
\end{figure*}

With a 1-bit error term, the maximum compression ratio is $32\times$. While this is already good, the compression ratio can be further improved by pruning the error term. Unstructured pruning usually has the best accuracy among pruning methods but often cannot result in a higher compression ratio since, in the worst case, 1 bit per weight is required to store the zero mask (worsening the compression ratio). Many structured pruning algorithms have been proposed, including CIM-aware versions to reduce the index (zero mask) overhead, with some slight accuracy trade-offs \cite{sie2022mars}. However, while these approaches achieve a higher compression ratio than unstructured pruning, special decoder circuits are still required to convert the index into zero masks. Moreover, since such algorithms are not fully structured, they often cannot result in an actual reduction in CIM size. This is because different filters or channels can have different masks (as shown in Figure \ref{fig:errorpruning} (a)). If the actual CIM cells are also pruned, this can lead to reduced parallelization and complicated input-switching logic.

CIMPool takes a pruning approach that minimizes the hardware overhead and lets the training algorithm solve the problem. In CIMPool, the error term pruning is fully structured. For 50\% sparsity, the first of every 2 channels has the error term, and the other one is pruned, while for 75\% sparsity, only the first of every 4 channels has the error term, as shown in Figure \ref{fig:errorpruning} (b) and (c). The same holds true for $87.5\%$ sparsity. This structure is fixed and shared with the entire network. Therefore, no index or zero mask is needed; only the value of the error term needs to be stored.

The structured pruning makes the effective error term storage scale with sparsity in a perfectly linear way, regardless of the layer dimension, unlike other pruning algorithms \cite{yue2020processorcim}. Also, the error array (as introduced in Section \ref{sec:architecture}) can be physically shrunk without extra hardware as the zero positions are fixed for all weight vectors.

\paragraph{Error scaling factor}

Enforcing structured pruning will naturally result in an accuracy drop, which could be significant  without optimization. To address it, we introduce an error scaling factor $S$, which can considerably reduce the accuracy drop. The idea is that as the error term becomes sparse, the values that remain carry more importance and should have a larger magnitude. For the case without error term pruning, the error term is scaled by the mean absolute value of the error. In this case, the scaling factor is multiplied on top of it, and the layer output is calculated as:
\begin{equation}
O = I \times MAV(W_{ori})\times W_{wp} + I \times S\times MAV(E)\times E_q
\end{equation}

, where MAV() stands for mean absolute value.

Table \ref{tab:errorscaleacc} shows the accuracy of ResNet-18 on Food-101 \cite{food101} for different error term sparsity and scaling factors (reason for choosing this dataset is explained in Section \ref{sec:eval}). The results demonstrate the significance of the error scaling factor, which could improve the accuracy by close to $10\%$ compared to the default setup.

\begin{table}[htb]
\caption{Accuracy on Food-101 dataset with different error scaling factor and error term sparsity.}
\label{tab:errorscaleacc}
\begin{tabularx}{\linewidth}{c|YYYY}
\hline
Scaling factor & 1       & 2       & 3       & 4       \\ \hline
0.5 sparsity      & 64.24\% & 67.19\% & 66.68\% &  64.08\%    \\
0.75 sparsity    & 55.92\% & 64.51\% & 65.79\% & 66.69\% \\ \hline
\end{tabularx}
\end{table}

\subsection{Network mapping}
When mapping the network to CIMPool hardware, the channel-first order is adopted, as the weight pool vectors correspond to the channel dimension as previously described. A single spatial position of the filters (XY dimension) will be mapped to the CIM at a given time, with input channels mapped to CIM rows and filters mapped to CIM columns. This mapping is also applied to the error array, and only the channels with error terms are mapped to the error pool (only they are stored in the error SRAM).

For the cases where the number of channels and number of filters are larger than the weigh pool array size, they are split into groups of 128 and multiple cycles are required to process them fully, similar to a conventional CIM array. When the channel size is smaller than the array size, zero will be fed into the extra CIM rows so that the result will not be impacted. 

\subsection{Overall framework flow} \label{sec:overallflow}
\begin{figure*}[ht]
\centering
\includegraphics[width=0.9\linewidth]{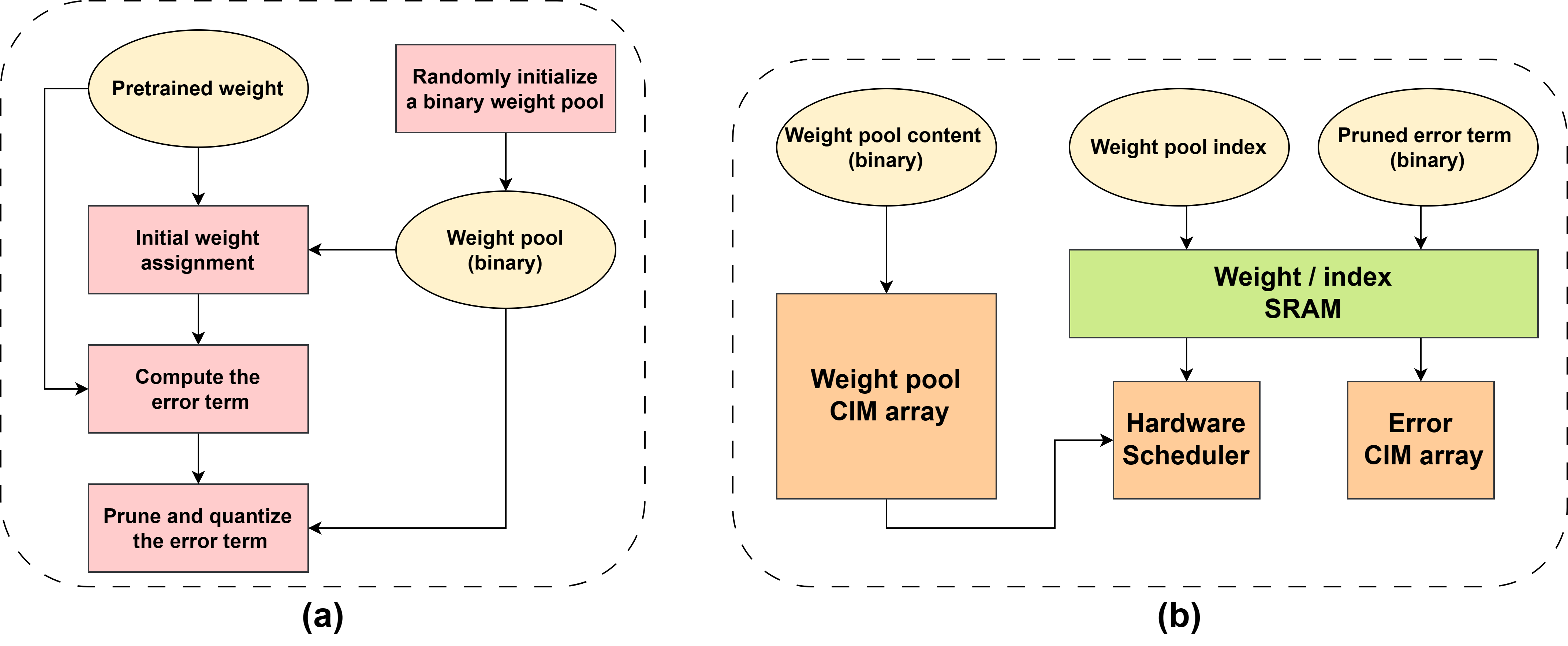}
\caption{(a): Training flow to generate the weight pool index and the error term. (b): Mapping generated weight pool content, index, and error term to CIMPool hardware.}
\label{fig:overallflow}
\end{figure*}

Figure \ref{fig:overallflow} illustrates the high-level flow of training and hardware mapping of CIMPool. At the beginning of the training phase, the model is loaded with a pretrained weight using full precision as a starting point. Then, during the retraining phase, the weights are first assigned to the vectors in the weight pool, and then the errors are calculated, pruned, and quantized. Afterward, the weights after the weight pool assignment and the errors are added together to form the actual weight being used by that layer. While the weights are assigned to the fixed weight pool vectors in the forward pass, the weights are still being trained during this process, and this change will be reflected in the error term.

Once the CIMPool training is completed, the weight pool indices and the errors are generated and stored in the weight and index SRAM while the weight pool contents are directly mapped into the weight pool CIM. The weight pool has dimension $(128,128)$ while the error terms' dimension either is the same or smaller than the original weight dimension, depending on sparsity. The indices are used in the hardware permutation unit to permute the output back to the original order.

\section{CIMPool Architecture} \label{sec:architecture}

\subsection{High-level architecture}

\begin{figure}[htbp]
\centering
\includegraphics[width=0.85\columnwidth]{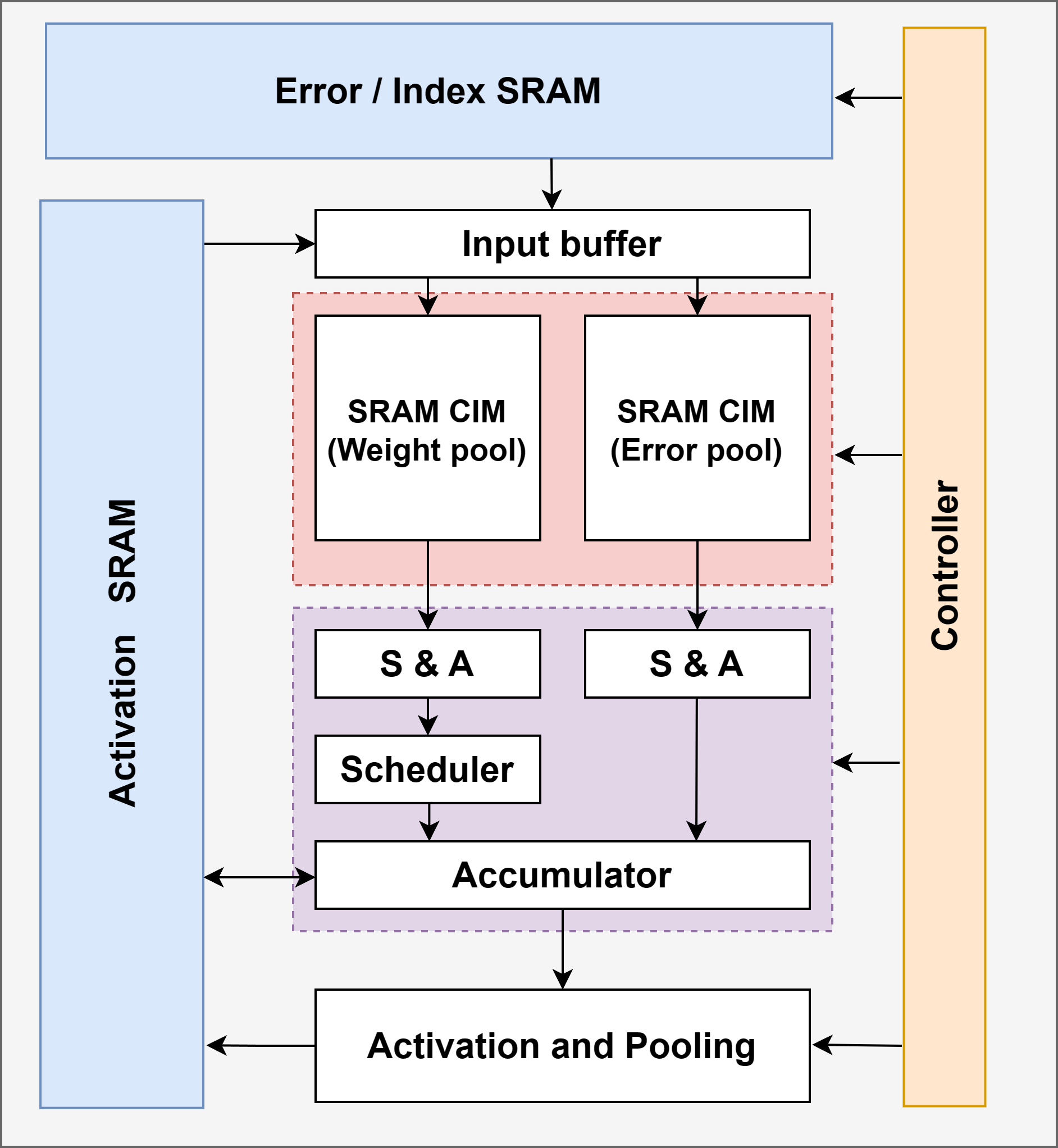}

\caption{High-level architecture diagram of CIMPool hardware, the diagram is not drawn to scale.}
\label{fig:hardwarehighlevel}
\end{figure}

As shown in Figure \ref{fig:hardwarehighlevel}, CIMPool only requires two CIM arrays for computation. The first CIM array is the weight pool array with a dimension of $128 \times 128$, which holds the contents of the entire weight pool. The content of the weight pool CIM array is fixed during the entire execution. 

The secondary CIM array is the error array, which holds the errors generated by the CIMPool algorithm. The error array's size can be $1-8 \times$ smaller than the weight pool array, depending on the target compression ratio. The error array can have a non-square shape, with the number of columns fixed to 128 and the number of rows ranging from 16 to 128. Unlike the weight pool array, the contents of the error array need to be updated during the execution, as errors are more like the weights of a conventional neural network. Updating contents of the error array has minimum latency overhead when weight-stationary dataflow is used, as SRAM allows fast memory writes, which is not possible for other types of CIMs. Weight-Stationary (WS) dataflow minimizes the frequency of updating the error pool contents, which is desired for CIMPool. WS dataflow can also reduce the overhead of hardware permutation, which will be discussed later in this section. 
For both the weight pool array and error array, the weight bitwidth is set to 1 to minimize the SRAM array area, and ADCs are not shared between columns as each column processes a unique filter.

Given the entire CIMPool hardware only contains two SRAM CIM arrays, the majority of the area can be allocated error/index memory. Hence larger models could be entirely stored on-chip without the need to access DRAM, which can be both a latency and energy bottleneck. The error/index memory is named this way because it stores the indices to the weight pool vectors as well as the computed errors. During the computation of each layer, the errors are fetched from the error memory and loaded into the error CIM array, while the input vectors are broadcasted to both the weight pool array and error array. 
Since the order of weight pool array outputs are permuted because of the weight pool assignment, a hardware scheduler is required to permute back the output to the correct order, which will be discussed in detail in Section \ref{sec:hwpermutation}. After permuting the output of weight pool CIM, the outputs of both arrays are digitally accumulated to form the correct partial sum. 

CIMPool rewrites the error array rather than fully unrolling it mainly for area efficiency, as storage SRAM arrays are typically much more compact than compute SRAM arrays (because of the extra circuitry for CIM computation). However, it is also possible to fully unroll the error CIM array just like conventional CIMs, without the need for hardware permutation or scheduling. 

\subsection{Hardware permutation} \label{sec:hwpermutation}
\subsubsection{Why hardware permutation is required}
\begin{figure}[htbp]
\centering
\includegraphics[width=0.95\columnwidth]{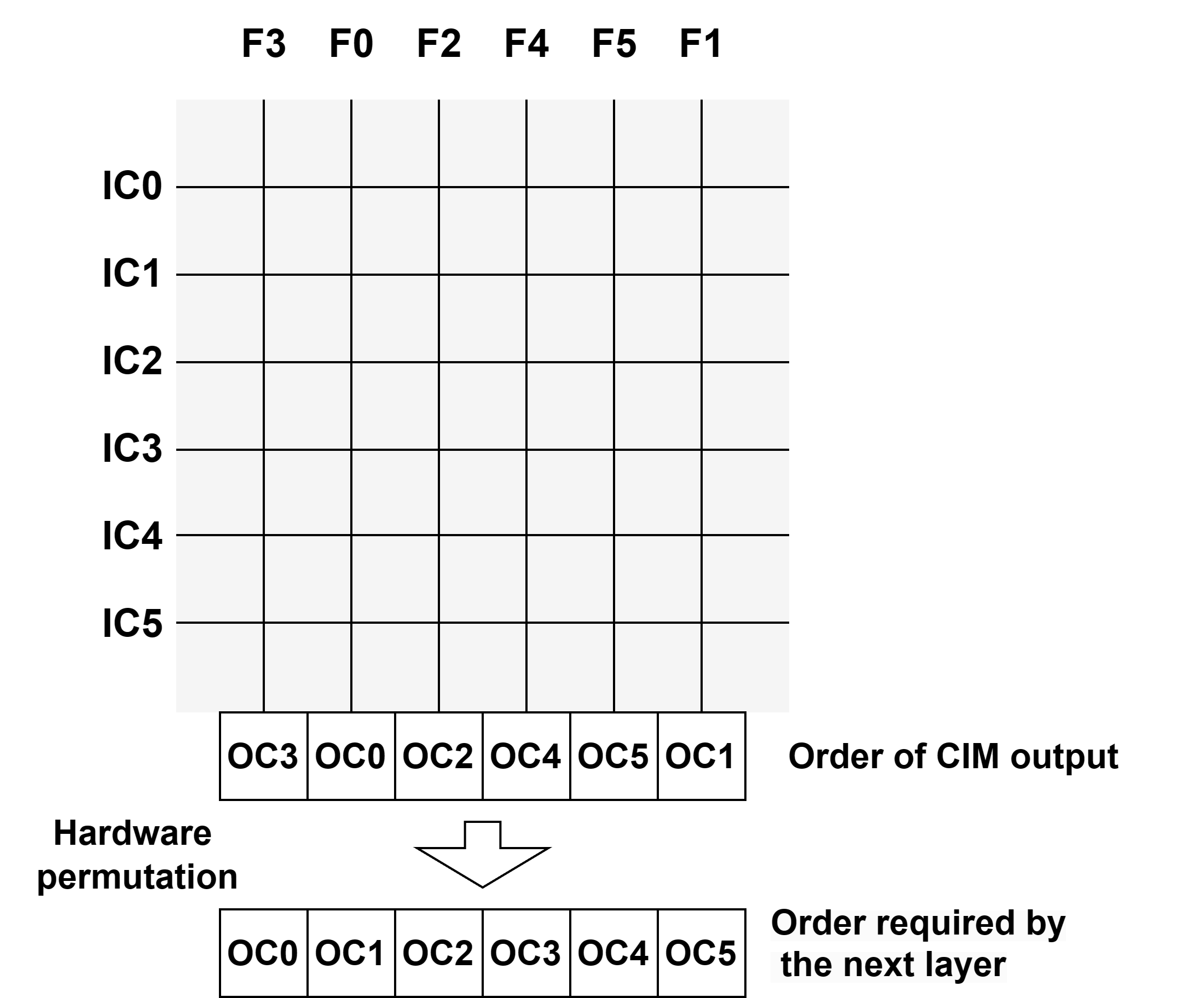}
\caption{Illustration of the necessity of hardware permutation. The CIM outputs need to be permuted back before being used as the inputs for the next layer since the input channels need to maintain the normal order.}
\label{fig:whypermute}
\end{figure}

Similar to conventional CIMs, since the inputs are broadcasted to the CIM columns, it's natural to let each CIM column process one unique convolution filter as different filters share the same input. After the weight pool assignment process of CIMPool, every filter that will be processed by the CIM at the same time is assigned to a unique weight vector in the weight pool (CIM column). Since the contents of the weight pool CIM are fixed, the order of CIM outputs depends on the weight pool assignment (which filter maps to which CIM column), as shown in Figure \ref{fig:whypermute}. 
In a typical neural network, the output channels of the current layer will be the input channels of the next layer. Because the CIM arrays can only take inputs in the normal channel order, the output of weight pool CIM needs to be permuted back to the original order before it is stored in the activation SRAM. 

\subsubsection{Mitigating the overhead} 
However, implementing the permutation in a hardware-efficient manner is not a trivial task - it could lead to significant overhead if not designed properly. Ideally, the permutation should be as fast as possible, such as within 1 cycle as a new set of CIM outputs could be produced each cycle. Implementing this in 1 cycle requires a $128 \times 128 $ crossbar with multi-bit channels, which is not feasible. Even if such a crossbar can be designed, the area and power overhead will overshadow all the benefits of CIM computation. 
A more power and area-efficient way to implement the permutation is to store the 128 outputs in a buffer, and read outputs from the buffer one by one according to the correct order. While this approach works and does not need crossbars, it can lead to significant latency and throughput overhead as it requires 128 cycles to perform the permutation. 

CIMPool leverages the property of generic CIM computation and optimizes the dataflow and weight pool design to implement hardware permutation efficiently without the need for crossbars. The hardware permutation logic in CIMPool, which is named hardware scheduler, can permute the entire output vector (128 values) at the same rate as input vectors are being processed by the CIMs on average while minimizing the latency and power overhead.

\paragraph{Parallel permutation}

The hardware scheduler in CIMPool is an optimized version of the second example discussed earlier - the crossbar-free version that requires 128 cycles to complete the permutation. Obviously, 128 cycles per one output vector won't work as the entire computation will be stalled due to the permutation process. One solution to this problem is permuting in parallel. The idea is since it takes 128 cycles to permute the output vector of the CIM array without a crossbar, a buffer can be designed to store all output vectors that can be produced in 128 cycles and perform the permutation in parallel. With WS dataflow, the indices of consecutive output vectors are the same as they share the same weight vectors. Therefore, the indices can be shared for the selectors (address decoders) in parallel.

\begin{figure*}[htp]
\centering
\includegraphics[width=0.95\linewidth]{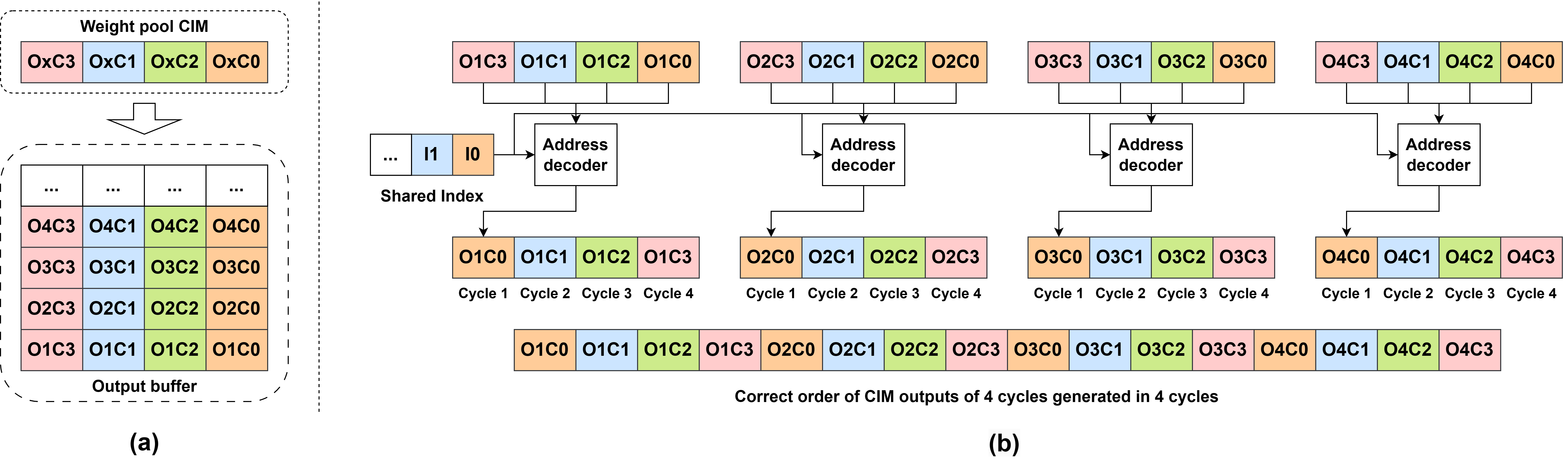}
\caption{Illustration of the parallel scheduling process used in the hardware scheduler, achieved by output buffers. Ix stands for Index for channel x, Ox stands for output vector x, and Cx stands for channel x.}
\label{fig:schedulerbuffer}
\end{figure*}

This process is illustrated in Figure \ref{fig:schedulerbuffer} with a simplified example assuming the CIM has only 4 columns (weight pool size of 4). In this case, an output buffer can be designed to hold 4 consecutive output vectors. 4 address decoders (selectors) are required with each one corresponding to one output vector, and the same index can be shared with all decoders as all output vectors have the same channel order. In each cycle, one output is read from each selector in parallel according to the index, and the correct order can be recovered in 4 cycles. Therefore, the hardware scheduler in this example can permute 4 output vectors per 4 cycles, which means 1 output vector per cycle on average. 

While this seems reasonable for the simplified example, the overhead is quite significant for CIMPool's 128-wide weight pool CIM. In this case, because of the ping-pong buffer, the output buffer needs to hold $2 \times 128 \times 128 = 32768$ bytes assuming 8-bit output (256 cycles of 128-byte outputs), which is not small. Moreover, that means 128 cycles are required just for buffering before any permutation can even be performed. This can cause significant throughput overhead. The area and throughput overhead can be further reduced through bit serial processing and weight pool optimization. 

\paragraph{Bit-serial computation}
For SRAM CIMs, inputs are typically processed in a bit-serial fashion and the CIM outputs for the bits are accumulated in the shift-and-add module within the SRAM CIM before sending out to the buffer. The filters stay the same within the bit serial cycles. For 8-bit activations, which is a common choice for many generic neural network accelerators, it is equivalent to the CIM producing one output per 8 cycles from the output buffer's perspective. By leveraging bit-serial processing, the output buffer size is reduced by $8 \times$ to 4096 bytes and the initial buffer filling overhead is reduced to 16 input cycles, where one input cycle is defined as the number of CIM bit-serial cycles required to process a full input. 

While the 16-input-cycle buffer filling overhead is a significant reduction over the previous 128-cycle overhead, it may still be non-negligible for certain cases. 
When both the network and input resolution (for CNNs) are small, for layers near the end of the network, the input activation size can be so small (such as $4 \times 4$) due to the pooling layers so that the 16-cycle buffer filing overhead still matters. Besides, the area of the 4 KB buffer is also not negligible when compared to the CIM array area. 

\begin{figure*}[htp]
\centering
\includegraphics[width=\linewidth]{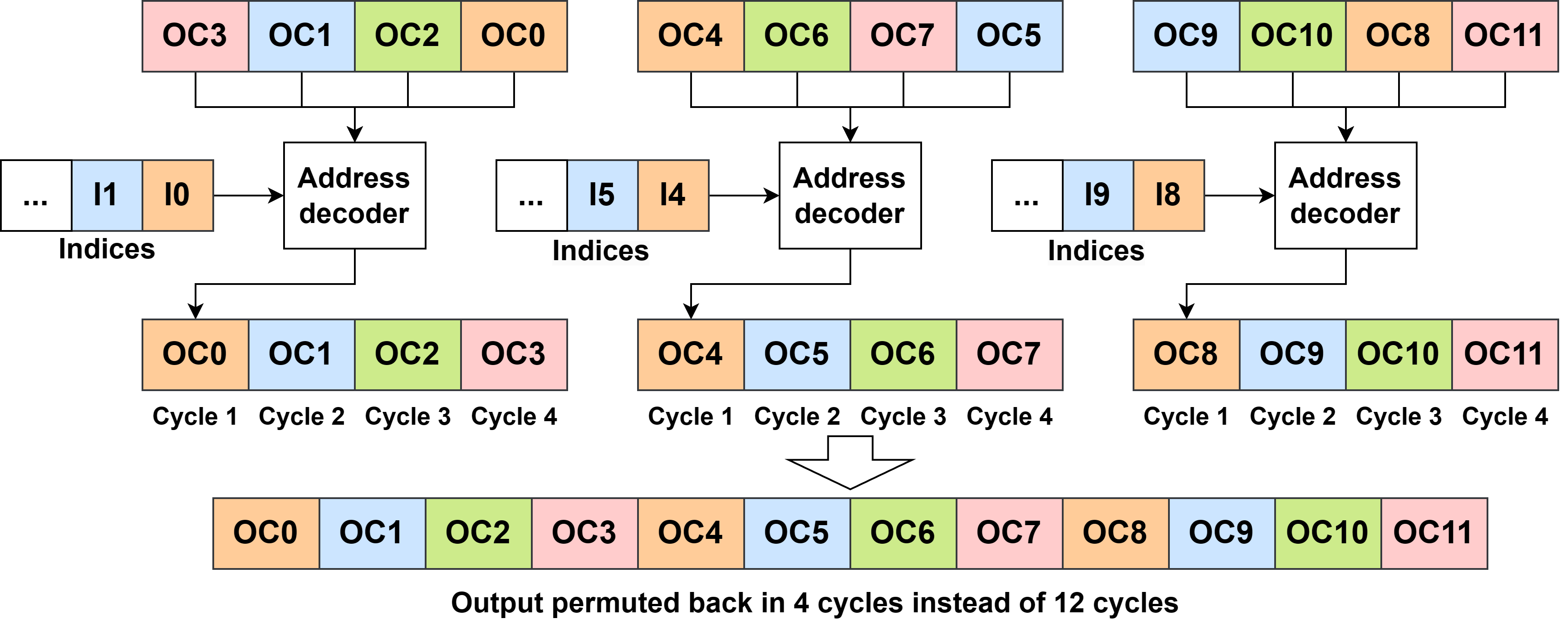}
\caption{Illustration of the grouped permutation of a single output vector with 12 channels. For simplicity, the circuits for other output vectors processed in parallel are not shown. The illustrated circuit is essentially one block of Figure \ref{fig:schedulerbuffer} (b). The abbreviations are the same as Figure \ref{fig:schedulerbuffer}.  }
\label{fig:schedulergroup}
\end{figure*}

\paragraph{Weight pool grouping}
One way to further reduce the 16-cycle overhead is through weight pool grouping, which means splitting the weight pool CIM into multiple independent groups and the original weight vectors can only select the vectors from the weight pool group it belongs to. For example, assume the original weight pool is split into 4 groups. Then each group contains 32 vectors of size $1 \times 128$. If filter 0-127 is scheduled to the weight pool CIM, then filter 0-31 can only be assigned to vectors in group 1, filter 32-63 can only be assigned to vectors in group 2, and so on. This approach reduces the effective weight pool size by $M$, where M is the number of weight pool groups.
The outputs of different groups are produced at the same time and so can naturally be permuted in parallel without the need for extra buffers. 
Figure \ref{fig:schedulergroup} illustrates an example with 12 CIM columns that are split into 3 groups. Without grouping, the 12 output values need 12 cycles to permute back to the original order. However, with grouping, the permutation of the outputs can happen in parallel among the three groups, and so only 4 cycles are required to complete the permutation.

The grouping can be applied on top of the existing buffered parallel permutation approach. Each selector block in Figure \ref{fig:schedulerbuffer} (b) can be replaced with the design shown in Figure \ref{fig:schedulergroup}. 

\section{Choosing the optimal group size for hardware scheduler} \label{sec:optimalgs}
\begin{figure}[htbp]
\centering
\includegraphics[width=\columnwidth]{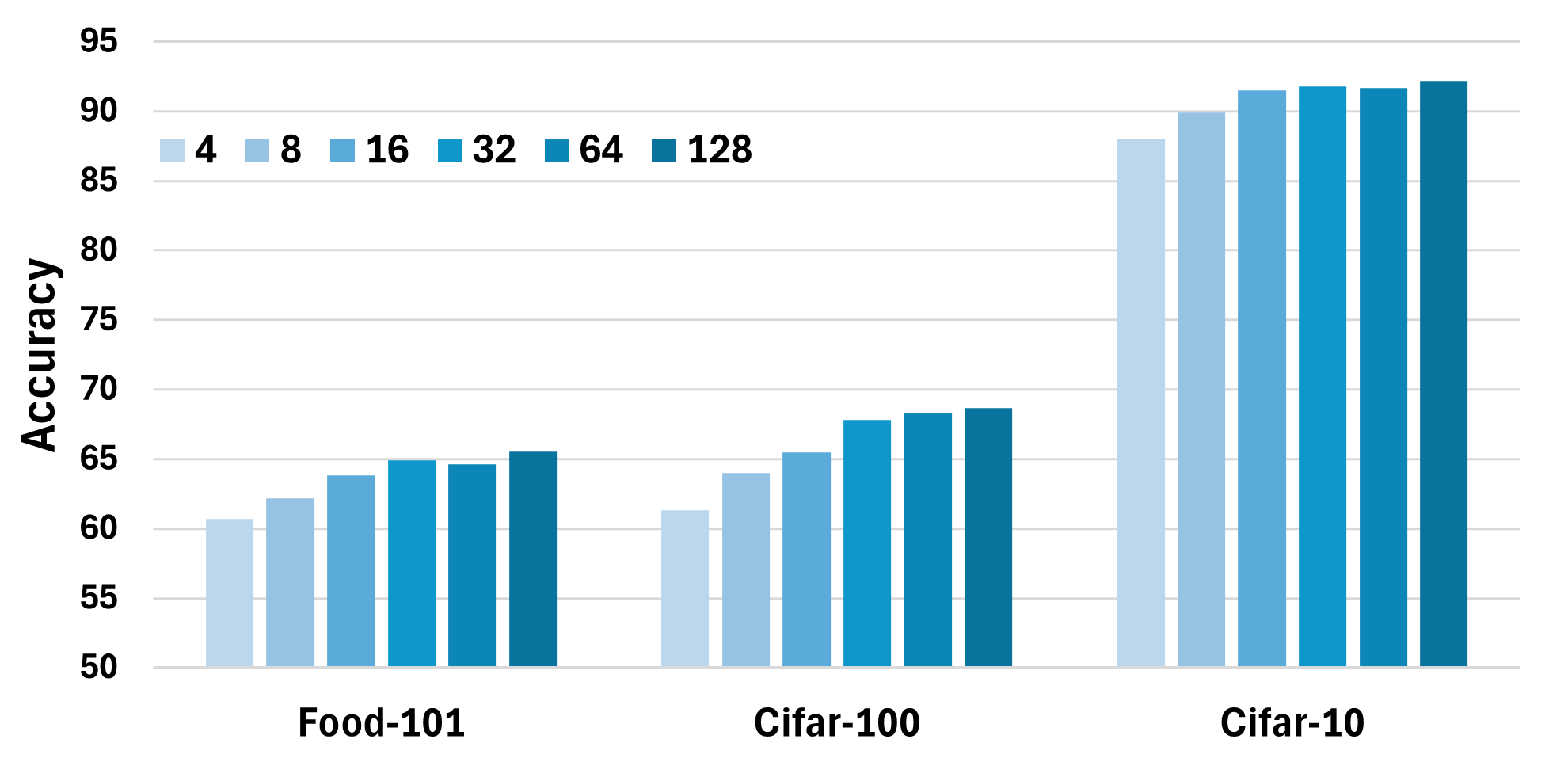}
\caption{Accuracy vs. weight pool group size results for ResNet-18 on three datasets. The error sparsity is set to 87.5\%. Group size of 128 is the case without grouping.}
\label{fig:accvsgroupsize}
\end{figure}

However, there are also costs for such improvement - the accuracy could be impacted as the effective weight pool size reduces. The impact is larger for an error sparsity value, as the network relies more on the weight pool in such cases. Therefore, to determine whether weight pool grouping works accuracy-wise and whether the optimal group size works, we compare the accuracy of different group sizes on three datasets (Food-101, CIFAR-100, and CIFAR-10) using ResNet-18. The error sparsity is set to 87.5\% to evaluate the impact of group size better. From the results shown in Figure \ref{fig:accvsgroupsize}, it's clear that a small group size such as 4 or 8 leads to a significant accuracy drop. This trend also demonstrates the importance of the weight pool, which is a crucial part of CIMPool. As group size increases, the accuracy drop is less significant. With a group size of 32, the accuracy is on par or close to the full weight-pool version (no grouping) on all three datasets. Therefore, we set the group size to 32 vectors for CIMPool, which achieves the optimal balance between accuracy and efficiency. With this group size, the weight pool CIM array contains 4 groups and hence reduces the output buffer storage requirement and initial buffer filling overhead by $4\times$. In this case, the initial buffer filling only requires 4 input cycles and the output buffer size is just 1024 bytes, which adds little overhead to the overall system.

Another advantage of weight pool grouping is the reduced index bits. For a full-weight pool with 128 vectors, a 7-bit index is required for each weight vector. However, with a group size of 32, the index bit is reduced to 5-bit, which further improves the compression ratio.

\section{Evaluation} \label{sec:eval}

We evaluate the accuracy results of CIMPool using ResNet-18 and ResNet-34 on three datasets: Food-101, CIFAR-10, and CIFAR-100. Food-101 and CIFAR-100 are chosen as they are more challenging than the CIFAR-10 dataset and are suitable to demonstrate CIMPool's capability of maintaining accuracy at a low compression ratio. Food-101 is a dataset with 101 categories of foods. The training data are not cleaned and contain a certain amount of noise that makes the dataset challenging.

We use PyTorch 2.1 to train all the models used in the evaluation, and we obtain the ResNet-18/34 implementation and the three datasets directly from Torchvision. We set the number of training epochs to 50 for all cases, and a best-effort approach is taken to tune the hyperparameters for all results, including comparison points. For Food-101, since the dataset contains inputs of various sizes, all inputs are resized to $256 \times 256$. All activation precision in accuracy evaluation is set to 8-bit, and an 8-bit ADC is assumed.

We generate all the hardware results using our simulator, which is heavily modified based on DNN+NeuralSim 2.1 \cite{dnnneuralsim}. Modifications are implemented to fit the CIMPool hardware architecture. We use CACTI \cite{muralimanohar2009cacti} to obtain SRAM area and energy, and 7 nm technology node is used for all hardware results.

\subsection{Compression ratio analysis}

In CIMPool, the raw weight values are not stored in the memory; instead, they are represented by the indices to the weight pool and the pruned errors. For each weight vector (128 values), the index is fixed to 5 bits because of weight pool grouping, while the error bits depend on the error sparsity. Since the errors are quantized to 1-bit, the number of error bits for a single weight vector is simply 128 multiplied by the sparsity. Table \ref{tab:compressionratio} lists the total bits per vector and effective compression ratio for the three error sparsities used in CIMPool. With 87.5\% sparsity, CIMPool can reduce the total storage by $48.76 \times$ compared to an 8-bit network.

\begin{table}[ht]
\caption{Total bits per vector and effective compression ratio for different error sparsity. The compression ratio is calculated against an 8-bit baseline.}
\begin{tabularx}{\columnwidth}{l|YYY}
\hline
Sparsity              & 0.5   & 0.75  & 0.875 \\ \hline
Total bits per vector & 69    & 37    & 21    \\
Compression ratio     & 14.84 & 27.68 & 48.76 \\ \hline
\end{tabularx}
\label{tab:compressionratio}
\end{table}

\subsection{Accuracy evaluation}

Table \ref{tab:acc_eval} presents the accuracy evaluation results of CIMPool compressed ResNet-18 and ResNet-34 on three datasets: Food-101, CIFAR-100, and CIFAR-10, with varying levels of error term sparsity. For comparison, the accuracy of the models quantized to 8, 4, and 1 bit is also provided.

On the Food-101 dataset, CIMPool with 0.5 and 0.75 error sparsity achieves accuracy comparable to the 8-bit baseline for both ResNet-18 and ResNet-34. Notably, the ResNet-34 model with 0.75 error sparsity (69.6\%) outperforms the 4-bit quantization (67.9\%), highlighting the effectiveness of CIMPool in maintaining high accuracy while achieving significant compression.

For CIFAR-100, the accuracy of the CIMPool baseline and the 0.5 error sparsity version is close to that of 4-bit quantization for both ResNet-18 and ResNet-34. The 0.75 error sparsity version of CIMPool achieves accuracy similar to 4-bit quantization on ResNet-34, while the accuracy is lower than 4-bit on the smaller ResNet-18 (but higher than the binary weight network).

For CIFAR-10, the difference between various compression/quantization configurations is smaller compared to the other two 'harder' datasets. CIMPool with 0.5 error sparsity can match the accuracy of 4-bit quantization for all cases, while CIMPool with 0.75 error sparsity is closer to 1-bit level accuracy. While the 0.875 sparsity version has lower accuracy than other versions on both networks, it can still achieve less than $2.3\%$ accuracy drop with a close to $50\times$ compression ratio.

\begin{figure}[htp]
\centering
\includegraphics[width=\columnwidth]{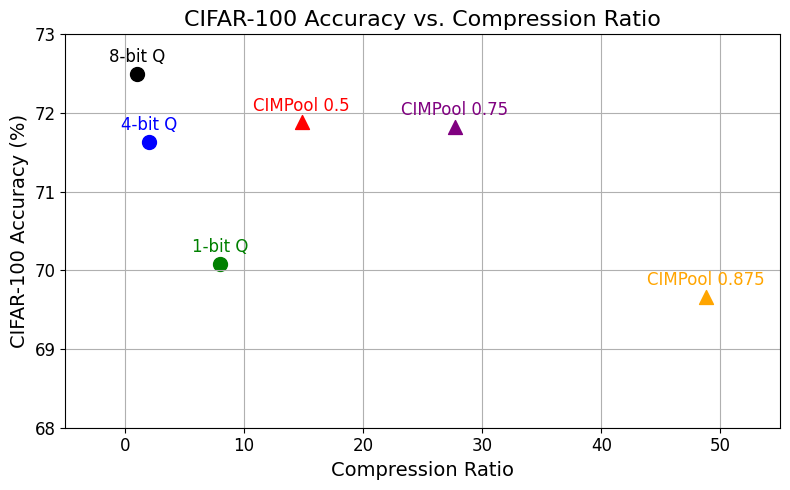}
\caption{Compression ratio versus accuracy on CIFAR-100, evaluated using ResNet-34 with different compression methods. Q stands for quantization and the number after CIMPool indicates its error sparsity.}
\label{fig:accvscompression}
\end{figure}

Figure \ref{fig:accvscompression} illustrates the accuracy of ResNet-34 on CIFAR-100 versus compression ratio for different compression and quantization methods. It is clear from the plot that CIMPool versions consistently achieve similar accuracy compared to quantization methods but have better compression ratios.

Based on the results, a good CIMPool setup is the 0.5 error sparsity version, which can achieve accuracy on par with the 8-bit baseline with $14.8 \times$ compression. Depending on the exact use case, CIMPool with 0.75 and 0.875 error sparsity can also be used with some accuracy and area trade-off. 

\begin{table}[t]
\caption{Accuracy difference and compression ratio (against 8-bit baseline) of CIMPool on three datasets, evaluated using ResNet-18 and ResNet-34. The accuracy of quantization with different bitwidths is also shown for comparison. The original accuracy for the two baselines is shown, and other accuracies are reported in accuracy drop for clarity. CR. stands for compression ratio and the number after CIMPool indicates its error sparsity.}
\label{tab:acc_eval}
\begin{tabularx}{\linewidth}{l|lYYY}
\hline
Version                 & CR. & Food-101 & CIFAR-100 & CIFAR-10 \\ \hline
\multicolumn{5}{c}{ResNet-18}  \\ \hline \hline
8-bit baseline    & 1x   & 67.4 \%   &72.1 \% &   94.0 \%   \\ 
4-bit        & 2x   & -1.7 \%   & -0.2 \%  &  -0.8 \%   \\
1-bit        & 8x   & -4.3\%    & -3.6 \% & -1.4 \%   \\ \hline
CIMPool 0.5 & 14.8x   & +0.1 \%   & -1.0 \%  & -0.6 \%   \\
CIMPool 0.75.  & 27.7x  & +0.1 \%    & -2.5 \% & -1.4\%    \\
CIMPool 0.875  & 48.8x  & -3.6 \%    & -5.6 \% & -2.2 \%    \\ \hline
\multicolumn{5}{c}{ResNet-34}  \\ \hline \hline
8-bit baseline    & 1x   & 69.5 \%   &72.5 \% &   94.0 \%   \\ 
4-bit        & 2x   &  -1.6 \%    &  -0.9 \%  &  -0.7 \%   \\
1-bit        & 8x   & -3.6 \%    & -2.4 \% & -1.0 \%   \\ \hline
CIMPool 0.5 & 14.8x   & +0.1 \%   & -0.6 \%  & -0.5 \%   \\
CIMPool 0.75.  & 27.7x  & -0.0 \%    & -0.7 \% & -1.5 \%    \\
CIMPool 0.875  & 48.8x  & -0.7 \%    & -2.8 \% & -2.3 \%    \\ \hline
\end{tabularx}
\end{table}

\subsection{System-level evaluation}

Since CIMPool is more of a CIM scaling method rather than a specific design, the system-level evaluation here only acts as an example to demonstrate the effectiveness of CIMPool. In theory, CIMPool can be applied to any SRAM CIM design regardless of the technology node and CIM array configuration.
For system-level evaluation, we use 0.5 error sparsity as the default CIMPool setup, as it achieves decent accuracy on all evaluated datasets and networks. The most suitable comparison target should be 4-bit quantization based on the accuracy results.

\subsubsection{Latency}
In this evaluation, all baseline CIM designs and CIMPool variants are configured to have the same throughput for more direct comparison, and the actual throughput for networks and datasets used in  evaluation are reported in Table \ref{tab:fps}. 

\begin{table}[h]
    \centering
    \caption{Throughput of CIMPool with different datasets and networks, in frames per second (FPS).}
    \begin{tabularx}{\columnwidth}{l|YY}
        \hline
        & ResNet-18 & ResNet-34 \\
        \hline
        CIFAR-10 & 1496.7 & 851.0 \\

        Food-101 & 374.2 & 212.8 \\
        \hline
    \end{tabularx}
    \label{tab:fps}
\end{table}

\subsubsection{Area}

Table \ref{tab:arearesult} shows the chip area breakdown of CIMPool and a 4-bit CIM when running ResNet-18. The data for CIMPool with 0.875 sparsity are also included for reference. The CIM array area for CIMPool contains the area of both the weight pool and error pool CIM arrays, as well as the hardware permutation logic.
The activation SRAM size is determined to hold the maximum intermediate activation (8-bit) of $256 \times 256$ input size. For ResNet-18, CIMPool 0.5  can achieve a 62.3\% area reduction over the 4-bit baseline, mainly contributed by the smaller weight SRAM area.

The last two rows show the exploration results of the maximum model parameters that can be stored on-chip for a 100 $mm^2$ area budget for these three setups. We assume the activation SRAM size remains the same for simplicity since the chip area will be dominated by the weight SRAM. With a 100 $mm^2$ chip area budget, CIMPool with 0.5 sparsity is capable of storing models with as large as 790 million parameters on-chip, which is close to $8$ times larger than the 4-bit CIM. The model parameters can be further increased to 2606 million when CIMPool with 0.875 sparsity is used. CIMPool with 0.875 sparsity is useful for tasks where a heavily compressed large model can outperform a lightly compressed small model. Take the Food-101 dataset as an example; CIMPool with 0.875 sparsity achieves higher accuracy on ResNet-34 than CIMPool with 0.5 sparsity on ResNet-18, even though the former actually has less weight storage than the latter.

\begin{table}[ht]
\caption{Area comparison and scaling exploration for two CIMPool versions and 4-bit quantization. For the comparison part, ResNet-18 is used for calculating the weight SRAM area. The unit for all area results is $mm^2$. }
\label{tab:arearesult}
\begin{tabularx}{\linewidth}{l|YYY}
\hline
                           & 4-bit  & CIMPool 0.5 & CIMPool 0.875  \\ \hline
CIM array area               & 0.3   & 0.2      &     0.2         \\
Activation SRAM area         & 3.6   & 3.6      &     3.6        \\
Weight SRAM area             & 9.9   & 1.4       &     0.4       \\
Total area                   & 13.8  & 5.2      &       4.2      \\ \hline
Max weight SRAM area         & 96.1  & 96.2     &     96.2       \\
Max \# of parameters (M) & 106.8  & 790.3        &  2605.9     \\ \hline
\end{tabularx}
\end{table}

\subsubsection{Power}

We also consider the case where an off-chip DRAM is available, so that the weights are loaded from DRAM to the CIM chip during execution. We set the on-chip weight SRAM size large enough such that all weights only need to load from DRAM once during the execution (no DRAM spill-fills). We observe that without compression, DRAM access energy can be a major energy consumer for applications with less weight reuse (e.g., smaller datasets such as CIFAR), even if energy-efficient HBM2 \cite{hbm2} DRAM (4pJ/bit access energy) is assumed.

Table \ref{tab:powertable} shows the power breakdown of different CIM versions when running ResNet-18 on the Food-101 and CIFAR-100 datasets. HBM2 is used for DRAM, and a batch size of 1 is assumed. The discrepancy between the two datasets is caused by their different input resolutions, hence different total computations and amounts of reuse. It's clear that for the 8-bit baseline, compute energy dominates when the input size is large (256$\times$ 256 for Food-101) and DRAM energy is more significant for smaller inputs. CIMPool can significantly reduce energy on both compute and DRAM access energy, because of its smaller CIM array size as well as less DRAM traffic. Therefore, in either case, CIMPool has a significant energy efficiency advantage compared to the 8-bit and 4-bit baselines. 

For both CIM versions benchmarked, because of the massive compression, the DRAM access energy no longer dominates (even smaller than SRAM and CIM energy). This results in up to a $3.24 \times$ reduction in overall energy in iso-accuracy comparison (4-bit versus CIMPool 0.5) and up to $4.55 \times$ if some accuracy degradation can be tolerated.

\begin{table}[htb]
\caption{Power breakdown of running ResNet-18 on different CIM versions, for Food-101 and CIFAR-100. The unit for all power results in the table is $\mu$ J.}
\begin{tabularx}{\columnwidth}{l|YYYY}
\hline
              & CIM   & SRAM  & DRAM  & Total \\ \hline
\multicolumn{5}{c}{Food-101}  \\ \hline \hline
8-bit         & 1813.6 & 106.5 & 351.8 & 2271.9 \\
4-bit         & 906.8  & 99.0  & 175.9 & 1181.7 \\
CIMPool 0.5   & 343.5  & 92.4  & 23.8  & 459.7 \\
CIMPool 0.875 & 259.1  & 91.6  & 7.2   & 357.9 \\ \hline
\multicolumn{5}{c}{CIFAR-100}  \\ \hline \hline
8-bit         & 453.2 & 38.0  & 351.8 & 843.0 \\
4-bit         & 226.7  & 30.4  & 175.9 & 433.0 \\
CIMPool 0.5   & 85.9  & 23.8  & 23.8  & 133.5 \\
CIMPool 0.875 & 64.8  & 23.1  & 7.2   & 95.1 \\ \hline
\end{tabularx}
\label{tab:powertable}
\end{table}

\section{Conclusion}

CIMPool presents a scalable and efficient solution to the memory bottleneck problem faced by SRAM CIM accelerators when executing large neural networks. By introducing a novel weight pool-based compression method and several hardware-aware optimizations, CIMPool achieves significant compression ratios while maintaining high accuracy. 

Empirical evaluations demonstrate CIMPool's effectiveness in reducing chip area and enabling the execution of much larger models within a given area budget compared to conventional quantization methods. CIMPool achieves an impressive 14.8  $\times$ compression ratio with accuracy on par with 8-bit baselines and outperforms state-of-the-art CIM-aware compression techniques. 
By considering the underlying hardware characteristics and constraints during the compression process, CIMPool offers a comprehensive solution that can be readily applied to various SRAM CIM designs, irrespective of technology node and array configuration.

Overall, CIMPool paves the way for executing much larger models on-chip, unlocking the full potential of CIM architectures. When compared to iso-accuracy 4-bit CIM, CIMPool can achieve a 62.3\% area reduction for a DRAM-less setup, or reduce the total chip energy by $3.24 \times$ when DRAM is used.
CIMPool can also be used for CIMs based on low-endurance non-volatile memories as the weight pools do not need to be overwritten, albeit at the cost of maintaining a small SRAM error CIM. We believe CIMPool represents an important step towards scalable and efficient neural network acceleration using CIMs.

\section{Ackonledgement}
We would like to thank Xinle Jiang for his work on developing the custom compute-in-memory (CIM) energy simulator based on NeuralSim, which was instrumental to our analysis. We also thank Ravit Sharma for his help in running several training experiments that supported our evaluation.

\bibliographystyle{IEEEtran}
\bibliography{main} 
\end{document}